\documentstyle [12pt,epsf]{article}

\input{epsf}
\begin{document}
\begin{titlepage}
\begin{center}

 \vspace{-0.1in}

{\large \bf Virtual Processes and Superradiance\\ in \\Spin-Boson Models}\\
 \vspace{.3in}{\large\em M.~Aparicio Alcalde,\footnotemark[1]
 R. Kullock\,\footnotemark[2]\,and N.~F.~Svaiter\,\footnotemark[3]}\\
\vspace{.2in}
 Centro Brasileiro de Pesquisas F\'{\i}sicas,\\
 Rua Dr. Xavier Sigaud 150,\\
 22290-180, Rio de Janeiro, RJ Brazil \\

\subsection*{\\Abstract}
\end{center}

\baselineskip .16in

We consider spin-boson models composed by a single bosonic mode
and an ensemble of $N$ identical two-level atoms. The situation
where the coupling between the bosonic mode and the atoms
generates real and virtual processes is studied, where the whole
system is in thermal equilibrium with a reservoir at temperature
$\beta^{-1}$. Phase transitions from ordinary fluorescence to
superradiant phase in three different models is investigated.
First a model where the coupling between the bosonic mode and the
$j-th$ atom is via the pseudo-spin operator $\sigma^{\,z}_{(j)}$
is studied. Second, we investigate the generalized Dicke model,
introducing different coupling constants between the single mode
bosonic field and the environment, $g_{1}$ and $g_{2}$ for
rotating and counter-rotating terms, respectively. Finally it is
considered a modified version of the generalized Dicke model with
intensity-dependent coupling in the rotating terms. In the first
model the zero mode contributes to render the canonical entropy a
negative quantity for low temperatures. The last two models
presents phase transitions, even when only Hamiltonian terms which
generates virtual processes are considered.

\vspace{0,1in}
PACS numbers: 42.50.Fx, 05.30.Jp, 73.43.Nq

\footnotetext[1]{e-mail: \,aparicio@cbpf.br}
\footnotetext[2]{e-mail:\,\,rkullock@cbpf.br}
\footnotetext[3]{e-mail:\,\,nfuxsvai@cbpf.br}

\end{titlepage}
\newpage\baselineskip .18in

\section{Introduction}

\quad $\,\,$The implementation of new techniques and ideas has
lead to increase the interest in spontaneous radiation and
collective effects of spin-boson models in free space and cavities
\cite{meschede} \cite{meystre} \cite{ brune} \cite{rbrandes}. This
has been a field of very attractive research, where experimental
and theoretical progress have emerged, which can be useful in
implementing concepts of quantum information and quantum
computation in real physical systems \cite{steane}. The purpose of
the present paper is to investigate different spin-boson systems,
and the physical consequences that follow from assuming first that
in interaction Hamiltonian of the models we are also considering
the counter-rotating terms. In the interaction picture these terms
describe non-resonant processes in which the two-level system and
the bosonic field are excited or de-excited simultaneously. Second
we study a interaction Hamiltonian that generates quite particular
virtual processes. We consider in different models the possibility
that virtual processes generate a quantum phase transition and
also a phase transition at finite temperature from ordinary
fluorescence to a superradiant phase characterized by the presence
of a condensate.

We consider three different models, in which we assume that a
single mode of a bosonic field interacts with an ensemble of $N$
identical two-level atoms. The whole system is in thermal
equilibrium with a reservoir at temperature $\beta^{-1}$. First,
we study a modified version of the model discussed by Chang and
Chakravarty, Legget and others \cite{chang} \cite{legget}
 \cite{benatti}, which has been used to analyze
dissipation in quantum computers \cite{parma} \cite{reina}. Owing
to the coupling between the two-level systems and a bosonic mode,
and assuming that its intensity is generic, the zero mode
contributes to render the canonical entropy a negative quantity
for low temperatures. After dealing with this problem, the
partition function is seen to be analytic for all temperature, and
therefore there is no phase transition in the model. Second, we
study the generalized Dicke model \cite{dicke} \cite{hioe}
\cite{carmichael} \cite{duncan}. We introduce different coupling
constants between the single mode of the bosonic field and the
ensemble of $N$ atoms, $g_{1}$ and $g_{2}$, for rotating and
counter-rotating terms, respectively. In a situation where only
virtual processes contribute, the generalized Dicke model present
a second order phase transition from the ordinary fluorescent to
the superradiant phase respectively, at some critical temperature
$\beta^{-1}_{c}$ \cite{tese} and also a superradiant phase
transition at zero temperature \cite{hertz} \cite{sachdev}, i.e.,
a quantum phase transition. We finally study the generalized Dicke
model where the coupling between the bosonic mode and $N$
two-level atoms is intensity-dependent \cite{jj1} \cite{jj2}
\cite{jj3}. The intensity-dependent contribution appears only in
the real processes. In the last model we are showing the presence
of the quantum phase transition.

The physics of superradiance is well known \cite{bonifacio1}
\cite{bonifacio2} \cite{yarunin} \cite{haroche} \cite{andreev}.
Let us suppose $N$ identical two-level atoms prepared in the
excited state. For dilute atomic systems, where there is no
coupling between the atoms, the $N$ atoms radiate incoherently,
where the radiation rate is proportional on the number of the
atoms $N$. Since the non-decay probability is exponential in time,
the intensity of emitted light has the time dependence
$I(\tau)=I_{0}\,e^{(-\frac{\tau}{\tau_{e}})}$, where $\tau_{e}$ is
the characteristic time for the spontaneous emission. Other
characteristic of the ordinary fluorescence is that the radiation
pattern of the atoms is isotropic. For an ensemble of atoms in a
volume small compared to the emission wavelength, they start to
radiate spontaneously much faster and stronger than the ordinary
fluorescence case, where the radiation rate becomes
quadratic-dependent on the number $N$ of atoms. Other important
characteristic of this cooperative process is that the emission
has a well defined direction depending upon the geometry of the
sample.

We are using path integral approach with the functional
integration method to investigate the thermodynamic of the models,
which is given by the analytic properties of the partition
function in the complex $\beta$ plane \cite{pop2} \cite{pri}
\cite{popov}. To study the nonanalytic behavior of the partition
function of the models using functional methods two steps are
mandatory. First, it is necessary to change the atomic pseudo-spin
operators of the models by a linear combination of Grassmann Fermi
fields to define fermionic models. Second, the thermodynamic limit
($N\rightarrow \infty$), where $N$ is the number of two-level
atoms must be taken . For the first two models that we are
interesting, the coupling between the pseudo-spin operators used
to describe the $N$ two-level atoms and the bosonic mode is
linear. Consequently the path integral describing the ensemble of
$N$ atoms is Gaussian and the integration over the degrees of
freedom describing the atoms can be performed exactly.

A summary of results obtained in the present literature in the
models is in order. The advantage of the first model is that it
allows an exact analytic solution, and presents destruction of
quantum coherence without decay of population. Concerning the
thermodynamics of the second spin-boson model, i.e., the Dicke
model, an important result was obtained by Hepp and Lieb
\cite{hepp}. These authors presented the free energy of the model
in the thermodynamic limit. For a sufficiently large value for the
coupling constant between the $N$ two-level atoms and the single
quantized mode of the bosonic field, there is a second order phase
transition from the normal to the superradiant phase. Later,
without assuming the rotating-wave approximation and by using a
coherent state representation, the study of the stability of the
model with an infinite number of bosonic modes was presented
\cite{hepp2}. The study of the phase transitions in the Dicke
model was presented also by Wang and Hioe \cite{wang}, where some
of the results obtained by Hepp and Lieb were reobtained. The
generalized Dicke model, where the counter-rotating terms are also
present in the interaction Hamiltonian, was investigated also by
Hioe \cite{hioe}, Carmichael et al. \cite{carmichael}, Duncan
\cite{duncan} and Li et al. \cite{li}. Hioe studied the
thermodynamics of the generalized Dicke model with two different
coupling constants using also the coherent states. Li et al.
pointed out that the $A^{2}$ term has been neglected in many
papers, since the presence of such term leads to non occurrence of
the quantum phase transition, as was proved by Rzazewski and
Wodkiewicz \cite{rz1} \cite{rz2}. Carmichael et al. claim that in
the limit $N\rightarrow\infty$, the thermodynamical properties of
the model are obtained by simply using the expression obtained
from the model with only the rotating terms and doubling the
coupling constant. As we will see, the thermodynamical properties
of the generalized Dicke model deserve a more careful analysis
\cite{tese}.

More recently  a bosonization procedure was employed to study the
phase transitions in the generalized Dicke model. Employing a
Holstein-Primakoff mapping \cite{prim} \cite{papanicolau}
\cite{hillery}, Emary and Brandes \cite{emary} \cite{emary2} were
able to express the generalized Dicke model in terms of a two
mode-bosonic field. These authors discussed the relation between
the quantum phase transition and the chaotic behavior that appear
in the model for finite $N$, where the energy level-spacing
statistics changes from Poissonian to one described by the
Gaussian ensemble of the random-matrix theory \cite{matrixmod}
\cite{braun}. This chaotic behavior was discussed also in the
Jaynes-Cummings model \cite{ja1} by Graham and Hoherbach
\cite{gra} \cite{gra2} and Lewenkopf et al. \cite{nemes}, in the
situation where the counter-rotating terms are present in the
interaction Hamiltonian \cite{muller} \cite{finney}, since the
seminal paper of Milonni et al. \cite{seminal}.

At this point, some remarks concerning the practical realization
of the generalized Dicke model in the laboratory is in order.
Dimer et al. \cite{dimer} pointed out that it remains a challenge
to provide a physical system where the counter-rotating terms are
dominant. In the Jaynes-Cummings model, Cirac et al. \cite{cirac}
discussed the possibility of controlling  the relative importance
of the counter-rotating terms using a ion trap. The question that
arises now is if there is some model where at some range of the
physical parameters, the counter-rotating terms are dominant. It
is possible to show that in the generalized Dicke model with
intensity-dependent coupling in the rotating terms the relative
importance of the contributions from the rotating and
counter-rotating terms can be controlled by changing the
temperature of the thermal bath. Concerning the
intensity-dependent coupling model, Buck and Sukumar \cite{jj1}
\cite{jj2} showed that the Heisenberg equations of motion can be
solved exactly and the behavior at finite temperature was
analyzed. Finally, Buzek \cite{jj3} using the rotating-wave
approximation also studied the same model, by presenting the time
evolution of the model.

This paper is organized as follows. In section II the path
integral with functional integral method is applied to study the
thermodynamics of a modified version of the model discussed by
Chang and Chakravarty, Legget and others.  In the section III we
study the thermodynamics of the generalized fermionic Dicke model.
In section IV we repeat our analysis for a generalized Dicke model
by including the intensity-dependent coupling in the terms
describing resonant processes. Conclusions are given in section V.
In this paper we use the terms environment and reservoir for a
system with a finite number of degrees of freedom and a system
with an infinite denumerable or not degrees of freedom,
respectively.  In the paper we use $k_{B}=c=\hbar=1$.

\section{The functional integral for the modified
fermionic Chang and Chakravarty model}

\quad $\,\,$ Let us consider a bosonic system $S$, with Hilbert
space ${\cal H}^{(S)}$ which is coupled with an ensemble of $N$
two-level atoms, with Hilbert space  ${\cal H}^{(B)}$. Let us
assume that the complete system is in thermal equilibrium with a
reservoir at temperature $\beta^{-1}$.
Let us denote by $H_{S}$, $H_{B}$ and $H_{I}$ are the Hamiltonians
of the bosonic field, the free $N$ two-level atoms, and the
interaction between both systems, respectively. The Hamiltonian
for the total system can be written as
\begin{equation}
H=H_{S}\,\otimes\,I_{B}+I_{S}\,\otimes\,H_{B}+H_{I}\, , \label{13}
\end{equation}
where $I_{S}$ and $I_{B}$ denotes the identities in the Hilbert
spaces of the bosonic field and the ensemble of $N$ atoms.

The aim of this section is to study the analytic behavior of
thermodynamical quantities, i.e., whether or not the system
exhibits a phase transition from normal to superradiant phase at
some critical temperature characterized by the presence of a
condensate in a model similar to the one introduced by Chang and
Chakravarty, Legget and others. Chang and others describes a
system of one two-level atom coupled to a reservoir of harmonic
oscillators, where the coupling between the reservoir and the atom
is done via the pseudo-spin operator $\sigma^{\,z}$. They studied
finite-time radiative processes in the model by assuming that the
environment is in thermal equilibrium. A generalization of this
model can be achieved introducing $N$ identical two-level atoms
and also an arbitrary mode-dependent coupling constant. Since
$[\,H,\sigma_{(j)}^{\,z}]=0$, this model allows for an exact
analytic solution. The Hamiltonian of this generalized model reads
\begin{equation}
 H=I_{R}\,\otimes\,\frac{\Omega}{2}\,
\sum_{j\,=1}^{N}\sigma^z_{(j\,)}+\sum_{k}\omega_{k}\,
a_{k}^{\dagger}\,a_{k}\,\otimes\,I_{S}+\frac{1}{\sqrt{N}}\sum_{j\,=1}^{N}
\sum_{k} \left(g_{j\,k}\,a_{k}^{\dagger}+
g_{j\,k}^{*}\,a_{k}\right)\,\otimes\,\sigma^{\,z}_{(j\,)},
\label{gen2}
\end{equation}
where $g_{jk}$ describes the coupling between the $j$-$th$
two-level atom with the reservoir. As usual, we are using the
pseudo-spin operators $\sigma_{(j)}^{+}$, $\sigma_{(j)}^{-}$ and
$\sigma_{(j)}^z$ which satisfy the standard angular momentum
commutation relations corresponding to spin $\frac{1}{2}$
operators. We are also shifting the zero of energy
$\frac{1}{2}(\omega_{1}+\omega_{2})$ for each atom and defining
$\Omega=\omega_{2}-\omega_{1}$.

 There are different physical situations that can be
investigated and they are characterized by the ratio of the
distances between the atoms $(r_{ij})$, which we assume to take
fixed positions, and the correlation length $L_{c}$ of the
reservoir, i.e., $\frac{r_{ij}}{L_{c}}<< 1$ and
$\frac{r_{ij}}{L_{c}}>> 1$. First, let us consider that the
minimal distance between the atoms is quite large if we compare it
to the correlation length of the reservoir. All the terms
$\frac{L_{c}}{r_{ij}}$ for all $i,j$ are nearly vanishing and
therefore each atom interacts with its own reservoir. Since the
phase transition from the fluorescent to the superradiant emission
is a cooperative process involving a collective mode of all the
atoms we concentrate our investigations in the limit where all the
terms $\frac{r_{ij}}{L_{c}}$ almost vanish. We consider the
situation where the linear dimension of the total atomic system is
small compared to the correlation length $L_{c}$ of the reservoir,
therefore the $N$ atoms interact collectively with the reservoir.
We can proceed by assuming that for a fixed $k$-mode of the
reservoir all the coupling constants $g_{ik}\,\,i=1,2,...,N$ of
Eq. (\ref{gen2}) are equal. The interaction between the reservoir
and the two-level atoms is given by the collective pseudo-spin
operators $J_{z}=\sum_{\,j=1}^{\,N}\,\sigma^z_{(j\,)}$. A
simplified model is achieved if we assume that the two-level atoms
and the bosonic modes are in the interior of a high-$Q$ cavity.
Assuming that the frequency $\omega_{0}$ of one of the cavity
modes is near-resonant with the energy gap $\Omega$ of the
two-level atoms, such as situation generates the following
physical model: the two-level atoms effectively interact only with
that mode, and all the other bosonic modes do not couple with the
two-level atoms. Under these circumstances the model is reduced to
a single mode of the bosonic field with the creation and
annihilation operators $b{^\dagger}$ and $b$, respectively,
interacting with an ensemble of atoms.

Let us define the Fermi raising and lowering operators
$\alpha^{\dagger}_{i}$, $\alpha_{i}$, $\beta^{\dagger}_{i}$ and
$\beta_{i}$, that satisfy the anti-commutator relations
$\alpha_{i}\alpha^{\dagger}_{j}+\alpha^{\dagger}_{j}\alpha_{i}
=\delta_{ij}$ and
$\beta_{i}\beta^{\dagger}_{j}+\beta^{\dagger}_{j}\beta_{i}
=\delta_{ij}$. We can also define the following bilinear
combination of Fermi operators, $\alpha^{\dagger}_{i}\alpha_{i}
-\beta^{\dagger}_{i}\beta_{i}$, $\alpha^{\dagger}_{i}\beta_{i}$
and finally $\beta^{\dagger}_{i}\alpha_{i}$. Note that
$\sigma^z_{(\,i)}$, $\sigma^+_{(\,i)}$ and $\sigma^-_{(\,i)}$ obey
the same commutation relations as the above bilinear combination
of Fermi operators. Therefore, we can change the pseudo-spin
operators of the spin-boson models by using the bilinear
combination of Grassmann Fermi fields

\begin{equation}
\sigma_{(i)}^{z}\longrightarrow (\alpha_{i}^{\dagger}\alpha_{i}
-\beta_{i}^{\dagger}\beta_{i})\, , \label{34}
\end{equation}
\begin{equation}
\sigma_{(i)}^{+}\longrightarrow \alpha_{i}^{\dagger}\beta_{i}\, ,
\label{35}
\end{equation}
and finally
\begin{equation}
\sigma_{(i)}^{-}\longrightarrow \beta_{i}^{\dagger}\alpha_{i}\, .
\label{36}
\end{equation}
Using the above results, the Euclidean fermionic action for the
model that we have defined above, can be written as
\begin{equation}
S = \int_0^{\beta} d\tau \left( b^*(\tau) \frac{\partial
b(\tau)}{\partial \tau} + \sum_{i=1}^{N} \left( \alpha_i^*(\tau)
\frac{\partial \alpha_i(\tau)}{\partial \tau} + \beta_i^*(\tau)
\frac{\partial \beta_i(\tau)}{\partial \tau} \right) \right) -
\int_0^{\beta} d\tau H_F(\tau), \label{s1}
\end{equation}
where $H_F$ is given by
\begin{eqnarray}
H_F=\omega_0\; b^*(\tau)b(\tau) + \frac{\Omega}{2}\sum_{i=1}^N
\Bigl(\alpha_i^*(\tau)\alpha_i(\tau) -\beta_i^*(\tau)\beta_i(\tau)\Bigr) \nonumber \\
+ \frac{g}{N}\sum_{i=1}^N
\left(\,\Bigl(\alpha_i^{*}(\tau)\alpha_i(\tau)-\beta_i^{*}(\tau)
\beta_i(\tau) \Bigr) \,\,\Bigl(b(\tau)+\,
b^{*}(\tau)\Bigr)\,\right)\,.
\end{eqnarray}
Usually, in spin-boson models, the coupling goes with
$\frac{1}{\sqrt{N}}$. Since this model has no crossed terms of
$\alpha_i$ and $\beta_i$, we are using a $\frac{1}{N}$ dependence
in the coupling constant. We interpret such change only as a
coupling constant renormalization. We can also justify the change
adopted using the Holstein-Primakoff mapping, where we obtain a
divergent Hamiltonian if we use the original coupling
$\frac{1}{\sqrt{N}}$. With the $\frac{1}{N}$ dependence in the
coupling constant the Hamiltonian of the system is well behaved.

Let us define the formal quotient of two functional integrals,
i.e., the partition function of the interacting model and the
partition function of the free model. Therefore, we are interested
in to calculate the following quantity
\begin{equation}
\frac{Z}{Z_{0}}=\frac{\int [d\eta]\,e^{\,S}}{\int
[d\eta]\,e^{\,S_{0}}}\, , \label{65}
\end{equation}
where $S=S(b,b^*,\alpha,\alpha^{\dagger},\beta,\beta^{\dagger})$
is the Euclidean action given by Eq. (\ref{s1}),
$S_0=S_{0}(b,b^*,\alpha,\alpha^{\dagger},\beta,\beta^{\dagger})$
is the free Euclidean action for the free single bosonic mode and
the free two-level atoms and finally $[d\eta]$ is the path
integral measure.
In Eq. (\ref{65}) we have functional integrals with respect to the
complex functions $b^*(\tau)$ and $b(\tau)$ and Grassmann Fermi
fields $\alpha_i^*(\tau)$, $\alpha_i(\tau)$, $\beta_i^*(\tau)$ and
$\beta_i(\tau)$. Since we are using thermal equilibrium boundary
conditions in the imaginary time formalism \cite{matsubara}
\cite{kubo} \cite{martin} \cite{fulling}, the integration
variables in Eq. (\ref{65}) obey periodic boundary conditions for
the Bose field, i.e., $b(\beta)=b(0)$, and anti-periodic boundary
conditions for the Grassmann Fermi fields i.e.,
$\alpha_i(\beta)=-\alpha_i(0)$ and $ \beta_i(\beta)=-\beta_i(0)$.

In order to obtain the effective action of the bosonic mode we
must integrate over the Grassmann Fermi fields. Therefore, let us
define the free action of the bosonic field by

\begin{equation}
S_{0}(b) = \int_{0}^{\beta} d\tau \biggl(b^{*}(\tau)
\frac{\partial b(\tau)}{\partial \tau} -
\omega_{0}\,b^{*}(\tau)b(\tau)\biggr)\,.
\end{equation}
The total action of the whole system can now be separated into
this free action and a Gaussian fermionic part. This is written in
the form

\begin{equation}
S = S_{0}(b) +  \int_{0}^{\beta} d\tau\,\sum_{i=1}^{N}\,
\rho^{\dagger}_{i}(\tau)\, M(b^{*},b)\,\rho_{i}(\tau)\, ,
\end{equation}
where $\rho_{\,i}(\tau)$ is a column matrix given in terms of
Grassmann  Fermi fields
\begin{eqnarray}
\rho_{\,i}(\tau) &=& \left(
\begin{array}{c}
\beta_{\,i}(\tau) \\
\alpha_{\,i}(\tau)
\end{array}
\right),
\nonumber\\
\rho^{\dagger}_{\,i}(\tau) &=& \left(
\begin{array}{cc}
\beta^{*}_{\,i}(\tau) & \alpha^{*}_{\,i}(\tau)
\end{array}
\right) \label{69a}
\end{eqnarray}
and the matrix $M(b^{*},b)$ is given by
\begin{equation}
M(b^{*}(\tau),b(\tau)) = \left( \begin{array}{cc}
\partial_{\tau} + \frac{\Omega}{2} + \frac{g}{N}(b(\tau)+\, b^{*}(\tau)) & 0 \\
0 & \partial_{\tau} - \frac{\Omega}{2} - \frac{g}{N}(b(\tau)+\,
b^{*}(\tau))
\end{array} \right)\, .
\label{matrix1}
\end{equation}
These complex functions and Grassmann Fermi fields $b(\tau)$,
$\alpha_{i}(\tau)$ and $\beta_{i}(\tau)$ can be represented in
terms of a Fourier expansion. Therefore, we have
\begin{equation}
b(\tau) = \beta^{-1/2} \sum_{\omega} b(\omega)\, e^{i\omega
\tau}\,
 \label{69c}
\end{equation}
and
\begin{equation}
\rho_{i}(\tau) = \beta^{-1/2} \sum_{p} \rho_{i}(p)\, e^{ip \tau}\,
. \label{71}
\end{equation}
Since the complex function $b(\tau)$ obeys periodic boundary
conditions, and the Grassmann Fermi fields $\alpha_{i}(\tau)$ and
$\beta_{i}(\tau)$ obey anti-periodic boundary conditions, we have
that $\omega = \frac{2\pi n}{\beta}$ and
$p=\frac{(2n+1)\pi}{\beta}$, where they are the bosonic and
fermionic Matsubara frequencies respectively.
Substituting the Fourier expansions given by the Eq. (\ref{69c})
and Eq. (\ref{71}) in the matrix $M_{p\,q}(b^{*},b)$ given by Eq.
(\ref{matrix1}) we get
\begin{equation}
M_{p\,q}(b^{*},b) = \left( \begin{array}{cc}
(ip + \frac{\Omega}{2})\delta_{p\,q} +  Q&0 \\
0 & (ip - \frac{\Omega}{2})\delta_{p\,q} - Q
\end{array} \right)\,
\label{73}
\end{equation}
where
\begin{equation}
Q = g\,N^{-1}\,\beta^{-1/2}\biggl(\,b^{*}(q-p) + b(p-q)\biggr).
\end{equation}
To simplify our calculations, let us change variables in the
following way:
\begin{equation}
b(\omega)\rightarrow~\Biggl(\frac{\pi}{(\omega_{0} -
i\omega)}\Biggr)^{1/2}b(\omega) \label{78}
\end{equation}
and
\begin{equation}
b^{*}(\omega)\rightarrow\Biggr(\frac{\pi}{(\omega_{0} -
i\omega)}\Biggr)^{1/2}b^{*}(\omega)\, . \label{79}
\end{equation}
Note that Eq. (\ref{79}) is not the complex conjugate of Eq.
(\ref{78}). It is easy to see that after these changes of
variables the denominator of the Eq. (\ref{65}) turns out to be
equal to unity
\begin{equation}
\int [d\eta(b)] \exp\biggl(-\pi
\sum_{\omega}b^{*}(\omega)b(\omega)\biggr) = 1\,. \label{80}
\end{equation}
We can express the ratio $\frac{Z}{Z_{0}}$ by the integral
\begin{equation}
\frac{Z}{Z_{0}}=\int[d\eta(b)]\,\exp{\biggl(\,S_{\,eff}\,(b)\,\biggr)}\,,
\label{81}
\end{equation}
where $S_{\,eff}(b)$ is the effective action of the bosonic mode
which is given by
\begin{equation}
S_{\,eff}\,=\,-\,\pi\,\sum_{\omega}\,b^{\,*}(\omega)\,
b(\omega)\,+\,N\ln{det\,(I\,+\,A)}\,. \label{81a}
\end{equation}
The matrix $A$ in the determinant of the above equation is given
by
\begin{equation}
(I+A) = \left( \begin{array}{cc}
I + E & 0 \\
0 & I + D
\end{array} \right)\,
\end{equation}
where the components of the matrices $E$ and $D$ are given by
\begin{equation}
E_{p\, q}=-\Biggl(\frac{\pi\,g}{\beta\,N}\Biggr)^
{\frac{1}{2}}\Biggl(ip-\frac{\Omega}{2}
\Biggr)^{-\frac{1}{2}}\Biggl(\,\frac{\,b^{*}\,(q-p)}{\sqrt{\omega_{0}-i(q-p)}}
+\frac{b\,(p-q)}{\sqrt{\omega_{0}-i(p-q)}}\Biggr)\,\Biggl(iq-\frac{\Omega}{2}
\Biggr)^{-\frac{1}{2}} \label{83c1}
\end{equation}
and
\begin{equation}
D_{p\, q}=\Biggl(\frac{\pi\,g}{\beta\,N}\Biggr)^{\frac{1}{2}}
\Biggl(ip+\frac{\Omega}{2}
\Biggr)^{-\frac{1}{2}}\Biggl(\,\frac{b^{*}\,(q-p)}{\sqrt{\omega_{0}-i(q-p)}}+
\frac{b\,(p-q)}{\sqrt{\omega_{0}-i(p-q)}}\Biggr)\,\Biggl(iq+\frac{\Omega}{2}
\Biggr)^{-\frac{1}{2}} . \label{83d1}
\end{equation}
In order to perform the functional integral given by Eq.
(\ref{81}) we must to find a manageable expression for
$det\,(I\,+\,A)$, so we use the following identity
\begin{eqnarray}
det\,(I\,+\,A)=det\,(I\,+\,E+\,D+E\,D)\simeq e^{tr\, E+tr\,
D-\frac{1}{2}E\,^2-\frac{1}{2}D\,^2} \label{det}
\end{eqnarray}
where the approximation in the exponent on the right-hand side of
Eq. (\ref{det}) is performed up to second order in the $E$ and $D$
matrices. Therefore, upon using the expressions of $E$ and $D$
given in Eq. (\ref{83c1}) and Eq. (\ref{83d1}), respectively, into
the approximation given in Eq. (\ref{det}), we can find an
expression for the ratio $\frac{Z}{Z_0}$ defined in Eq. (\ref{81})
of the following form
\begin{eqnarray}
\frac{Z}{Z_0} &=& \int[d\eta (b)] \exp \biggl(-\pi \sum_{\omega}
b^*(\omega)\,b(\omega)  + a_0\,\Bigl(b(0) + b^*(0)\Bigr)\nonumber\\
&-& \frac{1}{N}\sum_{\omega}c(\omega)\,\Bigl(b(\omega)\,b(-\omega)
+ b^*(\omega)\,b^*(-\omega)\Bigr) -
\frac{1}{N}\sum_{\omega}d(\omega)\,\Bigl(b(\omega)\,b^*(\omega)\Bigr)\biggr),
\end{eqnarray}
where the terms $a_0$, $c(\omega)$ and $d(\omega)$ are,
respectively, defined in the form

\begin{eqnarray}
a_0 &=&  g\Biggl( \frac{\pi\,\beta}{\omega_0}\Biggr)^{\frac{1}{2}}
\tanh{\left(\frac{\Omega\,
\beta}{4}\right)}, \\
c(\omega) &=& \frac{g^2\,\pi}{\beta (\omega_0^2 +
\omega^2)^{\frac{1}{2}}} \sum_p
\frac{\frac{\Omega^2}{4}-p(p-\omega)}{ \Bigl(p^2 +
\frac{\Omega^2}{4}
\Bigr)\Bigl((p-\omega)^2 + \frac{\Omega^2}{4}\Bigr)},\\
d(\omega) &=& \frac{2\,g^2\,\pi}{\beta (\omega_0 - i\omega)}
\sum_p \frac{\frac{\Omega^2}{4}-p(p-\omega)}{ \Bigl(p^2 +
\frac{\Omega^2}{4}\Bigr)\Bigl((p-\omega)^2 +
\frac{\Omega^2}{4}\Bigr)}.
\end{eqnarray}
It is possible to show that the contributions $c(\omega)$ and
$d(\omega)$ cancel out for $\omega\neq 0$ and $c(0)$ and $d(0)$
are different from zero, but their values do not matter in the
discussion next because in the limit $N\rightarrow\infty$ they do
not contribute to the partition function. Therefore in the
thermodynamic limit $N\rightarrow\infty$ we have that
$\frac{Z}{Z_0}$ can be written as
\begin{eqnarray}
\frac{Z}{Z_0} = \int\prod_{\omega\neq 0}db(\omega)\,db^*(\omega)\,
e^{-\pi \sum_{\omega\neq 0}  b^*(\omega)\,b(\omega)}\,\int
db(0)\,db^*(0)\, e^{-\pi\, b^*(0)\,b(0)+a_0\,\bigl(b(0) +
b^*(0)\bigr)}\,.
\end{eqnarray}
The integrations for $\omega\neq 0$ are not dependent on
thermodynamical parameters. Thus it can be considered as a
normalization constant, i.e.,
\begin{equation}
\frac{Z}{Z_0} = C_{0}\,\int db(0)\,db^*(0)\, e^{-\pi\,
b^*(0)\,b(0)+a_0\,\bigl(b(0) + b^*(0)\bigr)}\,.
\end{equation}
The partition function is therefore an entire function. From the
$\ln Z$ quantity, which is given by
\begin{eqnarray}
\ln Z=\ln Z_0 +\ln C_{0}
+\frac{g^2\,\beta}{\omega_0}\tanh^2\biggl(\frac{\Omega\,\beta}{4}\biggr)\,,
\label{zetap}
\end{eqnarray}
we can find the free energy of the model. Also the mean energy $E$
and canonical entropy $S$ can be computed. We get
\begin{equation}
E=E_0
-\frac{g^2}{2\,\omega_0}\tanh\biggl(\frac{\Omega\,\beta}{4}\biggr)\,\frac{\sinh
\Bigl(\frac{\Omega\,\beta}{2}\Bigr)+\Omega\,
\beta}{\cosh^2\Bigl(\frac{\Omega\,\beta}{4}\Bigr)}\,,
\end{equation}
and
\begin{equation}
S=S_0 + \ln C_{0}
-\frac{g^2\beta^2\Omega}{2\,\omega_0}\;\frac{\tanh\Bigl(\frac{\Omega\,\beta}{4}\Bigr)}
{\cosh^2\Bigl(\frac{\Omega\,\beta}{4}\Bigr)}\,
\end{equation}
where $S_{0}$ is the canonical entropy for the free case. Using
the third law of thermodynamics we get $C_{0}=1$.

We can see that the contribution of the last term in the
right-hand side of Eq. (\ref{zetap}) is coming from the zero mode.
Owing to the absence of the coupling via $\sigma_{(j)}^{+}$ and
$\sigma_{(j)}^{-}$ to the bosonic mode, the zero mode generates a
negative term in the expressions for the energy and the entropy.
It is known in the literature that the zero modes can lead to
problematic results \cite{dowker}. In the models discussed by
Dowker, the entropy becomes negative and also a
temperature-dependent pole appears in the free energy
\cite{dowker2}. At this point, as discussed before, there are two
different paths that we can follow concerning the zero-mode
problem. One is to disregard the zero mode, thus solving the
problem of the sign of the canonical entropy. We would like to
stress that we can not identify the zero-mode as a Nambu-Goldstone
mode  \cite{goldstone} \cite{goldstone2} since there is no
spontaneous breaking of the continuous symmetry in the model. The
second path is to assume that $g^{2}$ has some bound. For small
values of $g^{2}$ the model has positive entropy for finite
temperature and it goes to zero at $\beta\rightarrow \infty$,
satisfying the third law of thermodynamics. Both procedures, the
\textit{ad hoc} first one of throwing away the zero-mode, or
assuming some bound in the $g^{2}$ value, solve the problem of the
negative entropy, making the model consistent from the
thermodynamical point of view. Adopting any of these procedures,
the partition function is analytic for all temperatures.

\section{The functional integral for the generalized fermionic Dicke model}

\quad $\,\,$ The generalized Dicke model, where an ensemble of
identical $N$ atoms interacts linearly with one mode of a bosonic
field, is defined by the Hamiltonian
\begin{equation}
H= I_S\,\otimes\,\sum_{j=1}^{N}\,
\frac{\Omega}{2}\,\sigma_{(j)}^z+\omega_{0}\,b^{\dagger}\,b\,\otimes\,I_B+
\,\frac{g}{\sqrt{N}} \sum_{j=1}^{N}\,
\Bigl(b+b^{\dagger}\Bigr)\otimes
\Bigl(\sigma_{(j)}^{+}+\sigma_{(j)}^{-}\Bigr)\, . \label{i26}
\end{equation}
In the above equation $g$ is the coupling constant between the
atom and the single mode of the bosonic field. The $b$ and
$b^{\dagger}$ are the boson annihilation and creation operators of
mode excitations that satisfy the usual commutation relation
rules.

The aim of this section is to prove that a model with an
interaction Hamiltonian generating only virtual processes presents
a phase transition from normal to superradiant state at some
temperature with the presence of a condensate and also a quantum
phase transition at some critical coupling. Similarly to the last
section we find that the Euclidean action $S$ is given by Eq.
(\ref{s1}), where now $H_{F}$ is the full Hamiltonian for the
generalized fermionic Dicke model given by
\begin{eqnarray}
H_{F}\,=\,\omega_{0}\,b^{\,*}(\tau)\,b(\tau)\,+
\,\frac{\Omega}{2}\,\displaystyle\sum_{i\,=\,1}^{N}\,
\biggl(\alpha^{\,*}_{\,i}(\tau)\,\alpha_{\,i}(\tau)\,-
\,\beta^{\,*}_{\,i}(\tau)\beta_{\,i}(\tau)\biggr)\,+
\nonumber\\
+\,\frac{g_{\,1}}{\sqrt{N}}\,\displaystyle\sum_{i\,=\,1}^{N}\,
\biggl(\alpha^{\,*}_{\,i}(\tau)\,\beta_{\,i}(\tau)\,b(\tau)\,+
\alpha_{\,i}(\tau)\,\beta^{\,*}_{\,i}(\tau)\,b^{\,*}(\tau)\,\biggr)\,+
\nonumber\\
+\,\frac{g_{\,2}}{\sqrt{N}}\,\displaystyle\sum_{i\,=\,1}^{N}\,
\biggl(\alpha_{\,i}(\tau)\,\beta^{\,*}_{\,i}(\tau)\,b(\tau)\,+
\,\alpha^{\,*}_{\,i}(\tau)\,\beta_{\,i}(\tau)\,b^{\,*}(\tau)\biggr).
\label{66a}
\end{eqnarray}
Note we are introducing two coupling constants, $g_{1}$ and
$g_{2}$, for the rotating and counter-rotating terms,
respectively. As we discussed before, the main reason for this is
that we are interested in to identify the contribution of the real
and virtual processes in the phase transition with the formation
of the condensate.  We are interested in calculating the formal
quotient given by Eq. (\ref{65}). Following the same calculation
from last section we can arrive at Eq. (\ref{81}) and Eq.
(\ref{81a}), where the matrix $A$ is now given by
\begin{equation}
A\,=\,\left( \begin{array}{cc} 0 & B\\ -C & 0
\end{array} \right)\, .
\label{83}
\end{equation}
In the equation above the quantities $B$ and $C$ are matrices with
components given by
\begin{equation}
B_{p\, q}=\Biggl(\frac{\pi}{\beta
N}\Biggr)^{\frac{1}{2}}\Biggl(ip+\frac{\Omega}{2}
\Biggr)^{-\frac{1}{2}}\Biggl(\,\frac{g_{\,1}\,b^{*}\,(q-p)}{\sqrt{\omega_{0}-i(q-p)}}
+\frac{g_{\,2}\,b\,(p-q)}{\sqrt{\omega_{0}-i(p-q)}}\Biggr)\,\Biggl(iq-\frac{\Omega}{2}
\Biggr)^{-\frac{1}{2}} \label{83a}
\end{equation}
and
\begin{equation}
C_{p\, q}=-\,\Biggl(\frac{\pi}{\beta
N}\Biggr)^{\frac{1}{2}}\Biggl(ip-\frac{\Omega}{2}
\Biggr)^{-\frac{1}{2}}\Biggl(\,\frac{g_{\,1}\,b\,(p-q)}{\sqrt{\omega_{0}-i(p-q)}}
+\frac{g_{\,2}\,b^{*}\,(q-p)}{\sqrt{\omega_{0}-i(q-p)}}\Biggr)\,\Biggl(iq+\frac{\Omega}{2}
\Biggr)^{-\frac{1}{2}} . \label{83b}
\end{equation}
We shall investigate the integral given by Eq. (\ref{81}) for
temperatures that satisfy  $\beta^{-1}>\beta^{-1}_c$. First of all
let us show that this integral converges. Using the estimate
\begin{equation}
|\det(I+A)|\leq \exp\biggl(\mbox{Re}\,(tr A)+\frac{1}{2}\,tr (A
A^{\dagger})\biggr)\,,
\end{equation}
we can show that the ratio $\frac{Z}{Z_{0}}$ obeys the following
inequality
\begin{eqnarray}
\frac{Z}{Z_{0}}&\leq&\Biggl[\,\biggl(1\,-\,a_0(0)\,+
\,2\,c_0(0)\biggr)\,\biggl(1\,-\,a_0(0)\,-\,2\,c_0(0)\,\biggr)\,\Biggr]^{\,-\,1/2}\,
\nonumber\\
&&\prod_{\,\omega\,>\,0}\,\Biggl[\,\biggl(1\,-\,a_0(\omega)\,+
\,2\,c_0(\omega)\biggr)\,\biggl(1\,-\,a_0(\omega)\,-\,2\,c_0(\omega)\,\biggr)\,
\Biggr]^{\,-\,1}\,. \label{83c}
\end{eqnarray}
where the $a_{0}(\omega)$ and $c_{0}(\omega)$ are given
respectively by
\begin{equation}
a_{0}(\omega)\,=\,\frac{g_{\,1}^{\,2}\,+\,g_{\,2}^{\,2}}
{\beta\,(\omega^{\,2}_{\,0}\,
+\,\omega^{\,2})^{\,1/2}}\,\displaystyle\sum_{p\,-\,q\,=
\,\omega}\,\frac{1}{(\frac{\Omega^{\,2}}{4}\,+\,
q^{\,2})^{\,1/2}}\,\frac{1}{(\frac{\Omega^{\,2}}{4}\,+\,
p^{\,2})^{\,1/2}},
\end{equation}
and
\begin{equation}
c_{0}(\omega)\,=\,\frac{\omega_0\,g_{\,1}\,g_{\,2}}
{\beta\,(\omega^{\,2}_{\,0}\,
+\,\omega^{\,2})}\,\displaystyle\sum_{p\,-\,q\,
=\,\omega}\,\frac{1}{(\frac{\Omega^{\,2}}{4}\,+\,
q^{\,2})^{\,1/2}}\,\frac{1}{(\frac{\Omega^{\,2}}{4}\,+\,
p^{\,2})^{\,1/2}}.
\end{equation}
In a similar way like Popov and Fedotov \cite{popov} proved, for
the case of rotating wave approximation we have that
$0<a_0(\omega)\,+ \,2\,c_0(\omega)<a_0(0)\,+\,2\,c_0(0)$ and
$a_0(0)\,+\,2\,c_0(0)=O(\omega^{-2}\ln\,\omega)$. Therefore if
$a_0(0)\,+\,2\,c_0(0)<1$, then Eq. (\ref{83c}) guarantees
convergence of the expression $\frac{Z}{Z_0}$. The condition
$a_0(0)\,+\,2\,c_0(0)=1$ is the equation for the transition
temperature, then we have
\begin{equation}
a_0(0)\,+\,2\,c_0(0)\,=\,\frac{(\,g_{\,1}+g_{\,2}\,)^{\,2}}
{\Omega\,\omega_{0}}\,\tanh\biggl(\frac{\beta_{c}\,\Omega}{4}\biggr)\,=1\,.
\end{equation}
The inverse of the critical temperature $\beta_{c}$ is given by
\begin{equation}
\beta_{c} =
\frac{4}{\Omega}\tanh^{-1}\biggl(\frac{\Omega\,\omega_{0}}{(\,g_{\,1}+
g_{\,2}\,)^{\,2}}\biggr)\,.
 \label{acre}
\end{equation}
Note that there is a quantum phase transition where the coupling
constants $g_{1}$ and $g_{2}$ satisfy
$g_{1}+g_{2}=(\omega_{0}\,\Omega)^{\frac{1}{2}}$. For
$g_{1}+g_{2}\neq(\omega_{0}\,\Omega)^{\frac{1}{2}}$ the partition
function is no more an entire function in the positive half of the
complex plane for the temperature $\beta_{c}^{-1}$ given by Eq.
(\ref{acre}). The system enters in a superradiant phase. To
calculate the asymptotic behavior of the functional integrals at
temperatures that satisfy $\beta^{-1}>\beta^{-1}_{c}$, we can do
the following approximation
\begin{equation}
\det\,^{N}(I+A) = \det\,^{N}(I+BC)\rightarrow
\exp\biggl(N\,tr(BC)\biggr)\, . \label{84}
\end{equation}
After some calculations $\frac{Z}{Z_{\,0}}$ can be written as
\begin{eqnarray}
\frac{Z}{Z_{\,0}}&=& \Biggl[\,\biggl(1\,-\,a(0)\,+
\,2\,c(0)\biggr)\,\biggl(1\,-\,a(0)\,-\,2\,c(0)\,\biggr)\,\Biggr]^{\,-\,1/2}\,
\nonumber\\
&&\prod_{\,\omega\,>\,0}\,\Biggl[\,\biggl(1\,-\,a(\omega)\,\biggr)\,
\biggl(1\,-\,a(\,-\omega)\,\biggr)\,-\,c^{\,2}(\omega)\,\Biggr]^{\,-\,1}\,+
\nonumber
\\
\nonumber
\\
&&+O(N^{\,-1})\,, \label{95}
\end{eqnarray}
where $a(\omega)$ and $c(\omega)$ in the above equation are given,
respectively, by
\begin{equation}
a(\omega)\,=\,\Biggl(\frac{g_{\,1}^{\,2}\,(\Omega
-i\omega)^{\,-1}+\,g_{\,2}^{\,2}\,(\Omega\,+\,
i\omega)^{\,-1}}{(\omega_{0}\,-\,i\,\omega)}
\Biggr)\,\tanh{\biggl(\,\frac{\beta\,\Omega}{4}\,\biggr)}\,
\label{90}
\end{equation}
and
\begin{equation}
c(\omega)\,=\,\Biggl(\frac{g_{\,1}\,g_{\,2}\,\Omega}{(\omega_{\,0}^{\,2}\,+
\,\omega^{\,2})^{\,1/2}\,(\Omega^{\,2}\,+\,\omega^{\,2})}\Biggr)
\,\tanh{\,\biggl(\frac{\beta\,\Omega}{4}\biggr)}.
 \label{90a}
\end{equation}
Taking the limit ($N\rightarrow \infty$) in Eq. (\ref{95}) we are
in the thermodynamic limit. We turn out to the discussion
concerning the local elementary excitation of the ground state. To
find the collective excitation energy level spectrum we have to
use the equation
\begin{equation}
c^{\,2}(\omega)\,-\,\biggl(1\,-\,a(\omega)\,\biggr)\,
\biggl(1\,-\,a(\,-\omega)\,\biggr)\,=0\, , \label{105}
\end{equation}
and making the analytic continuation $(i\omega \rightarrow E)$, we
obtain the following equation
\begin{eqnarray}
&&1\,=\,-\Biggl[\frac{g_{\,1}^{\,4}\,+\,g_{\,2}^{\,4}}
{(\omega_{\,0}^{\,2}\,-\,E^{\,2})\,(\Omega^{\,2}\,-\,E^{\,2})}\Biggr]\,
\tanh^{\,2}\biggl(\frac{\beta\,\Omega}{4}\biggr)\,+
\nonumber\\
\nonumber\\
&&-\Biggl[\frac{g_{\,1}^{\,2}\,g_{\,2}^{\,2}}{(\omega_{\,0}^{\,2}
\,-\,E^{\,2})}\Biggl(\frac{1}{(\Omega\,-E)^{\,2}}\,+
\,\frac{1}{(\Omega\,+\,E)^{\,2}}\,-\,\frac{4\,\Omega^{\,2}}
{(\Omega^{\,2}\,-\,E^{\,2})^{\,2}}\Biggr)\,\Biggr]\,
\tanh^{\,2}\Biggl(\frac{\beta\,\Omega}{4}\Biggr)\,+
\nonumber\\
\nonumber\\
&&+\,\Biggl[\frac{g_{\,1}^{\,2}(\,\Omega\,-\,E\,)^{\,-1}\,+
\,g_{\,2}^{\,2}(\,\Omega\,+\,E\,)^{\,-1}}{(\omega_{0}\,-\,E)}\,+
\frac{g_{\,1}^{\,2}(\,\Omega\,+\,E\,)^{\,-1}\,+
\,g_{\,2}^{\,2}(\,\Omega\,-\,E\,)^{\,-1}}{(\omega_{0}\,+\,E)}\Biggr]
\,\tanh\Biggl(\frac{\beta\,\Omega}{4}\Biggr).\nonumber\\
\label{105}
\end{eqnarray}
Solving the above equation for the case $\beta^{-1}=\beta^{-1}_c$
we find the following roots
\begin{equation}
E_{\,1}\,=\,0
\label{106}
\end{equation}
and
\begin{equation}
E_{\,2}\,=\,\Biggl(\,\frac{g_{\,1}\,(\Omega\,+\,\omega_{\,0})^{\,2}\,+\,
g_{\,2}\,(\Omega\,-\,\omega_{\,0})^{\,2}}{(g_{\,1}\,+\,g_{\,2})}\,\Biggr)^{\,1/2}\,.
\label{107}
\end{equation}
Its low energy state of excitation is a Nambu-Goldstone mode. Now,
let us present the critical temperature and the energy level
spectrum of the collective bosonic excitations of the model with
the rotating-wave approximation, where $g_{1}\neq 0$ and
$g_{2}=0$.
The result obtained by Popov and Fedotov is recovered, where the
equation
\begin{equation}
a(0) = 1\, \label{102}
\end{equation}
and
\begin{equation}
\frac{g_{1}^{2}}{\omega_{0}\Omega}
\tanh\biggl(\frac{\beta_{c}\,\Omega}{4}\biggr) = 1\,, \label{103}
\end{equation}
give the inverse of the critical temperature, $\beta_{c}$. It is
given by
\begin{equation}
\beta_{c} = \frac{4}{\Omega}\,arctanh
\biggl(\frac{\omega_{0}\Omega}{g_{1}^{2}}\biggr)\, . \label{104}
\end{equation}
The order parameter of the transition is the expectation value of
the number of excitation associated to the bosonic mode per atom,
i.e., $lim_{N\,\rightarrow\infty}\,\frac{\langle\,b^
{\dagger}\,b\rangle}{N}\neq\,0$. Note that again $\omega_{0}$,
$\Omega$ and $g_{1}$ define also a non-zero critical temperature
where the partition function is no more an entire function in the
positive half of the complex $\beta$-plane. We may expect a
superradiant phase for the temperature $\beta_{c}^{-1}$ given by
Eq. (\ref{104}). The energy level spectrum of the collective Bose
excitations in this case is
\begin{equation}
E_{1}=0\, , \label{108}
\end{equation}
and
\begin{equation}
E_{2}=\Omega+\omega_{0}\, . \label{109}
\end{equation}
In this case, there is also a quantum phase transition, i.e., a
zero temperature phase transition when
$g_{1}=(\omega_{0}\,\Omega)^{\frac{1}{\,2}}$. Now we will show
that is possible to have a condensate with superradiance in a
system of $N$ two-level atoms coupled with one mode of a Bose
field where only virtual processes contribute.
In the pure counter-rotating wave case, i.e., $g_{1}=0$ and
$g_{2}\neq 0$, the inverse of the critical temperature,
$\beta_{c}$ is given by
\begin{equation}
\beta_{c} = \frac{4}{\Omega}\,arctanh
\biggl(\frac{\omega_{0}\Omega}{g_{2}^{2}}\biggr)\, , \label{112}
\end{equation}
and the spectrum of the collective Bose excitations given by
\begin{equation}
E_{1}=0\, , \label{110}
\end{equation}
and
\begin{equation}
E_{2}=|\,\Omega-\omega_{0}|\, . \label{111}
\end{equation}
A comment is in order concerning the spectrum of the Bose
excitations. In both of the cases: working with the pure
counter-rotating or the rotating-wave terms, there is a phase
transition. In the case of the rotating-wave approximation
$g_{1}\neq 0$ and $g_{2}=0$, there is a Nambu-Goldstone mode
$(E=0)$. In the pure counter-rotating case $g_{1}=0$ and
$g_{2}\neq 0$ also there is a Nambu-Goldstone (gapless) mode.
Thence we show that it is possible to have a condensate with
superradiance in a system of $N$ two-level atoms coupled with one
mode of a Bose field where only virtual processes contribute.
Since the energy $E_{2}\approx 0$, local elementary excitations of
the ground state with low energy can easily be created causing a
significant fluctuation effect. Unfortunately we are not able to
evaluate these effect in the systems.

An important question is the way of practical realization of the
second model, i.e., the generalized Dicke model in the laboratory.
As was stressed by Dimer et al \cite{dimer} it remains a challenge
to provide a physical system where the counter-rotating terms are
dominant.  Experimental observation of the superradiant phase in a
situation where is possible to control the importance of the
counter-rotating terms in the generalized Dicke model
($g_{1}\approx 0$, $g_{2}\neq\,0$) could improve our understanding
of this phenomenon.

\section{The functional integral for the fermionic
Dicke model with intensity-dependent coupling}
\quad $\,\,$ A model where the behavior is quite interesting from
the physical point of view is the one where the coupling between
$N$ identical two-level atoms and one mode of a bosonic field in a
lossless cavity is intensity-dependent. The generalization for
this model introducing the counter-rotating terms is
straightforward. Therefore, in this paper we discuss the model
with rotating and counter-rotating terms, where the
intensity-dependent contribution appears only in the real
processes. The necessity for disregard the intensity dependent
coupling contribution in the part of the interaction Hamiltonian
that generates the virtual processes is easy to justify. It is
counterintuitive that the virtual processes are amplified when the
number of excitations in the bosonic sector is enhanced. We will
show that at low temperatures the contribution from the
counter-rotating terms dominate over the rotating ones and the
system presents a quantum phase transition.

After this discussion, the total Hamiltonian of the atoms and the
bosonic mode takes the form
\begin{eqnarray}
H&=&I_S\,\otimes\,\sum_{j=1}^{N}\,\frac{\Omega}{2}\,\sigma_{(j)}^z+
\omega_{0}\,b^{\dagger}\,b\,\otimes\,I_B+
\nonumber\\
&&\frac{g}{\sqrt{N}}\sum_{j=1}^{N}\,
\Bigl(b\,(b^{\dagger}\,b)^{\frac{1}{\,2}}\otimes\,
\sigma_{(j)}^{+}\,+b^{\dagger}\, (b^{\dagger}\,b)^{\frac{1}{\,2}}
\otimes\,\sigma_{(j)}^{-}+b^{\dagger} \otimes\,
\sigma_{(j)}^{+}\,+b\, \otimes\,\sigma_{(j)}^{-}\Bigr)\, .
\label{aa28}
\end{eqnarray}
 At this point, let us make a parallel between the model given by
Eq. (\ref{aa28}) assuming the rotating-wave approximation, and the
Jaynes-Cummings model \cite{ja1}, where one two-level atom is
coupled to a single mode quantized electromagnetic field. It is
possible to generalize this model, by using the rotating-wave
approximation, known as the two-photon Jaynes-Cummings model
\cite{puri} \cite{toor}. The introduction of the counter-rotating
terms is straightforward. It has been discussed in the literature
the possibility of controlling the relative importance of them in
laboratory using an ion trap \cite{cirac}.

Going back to the model we have also non-Gaussian functional
integrals to solve. The main point we wish to demonstrate in this
section is that the model defined by the equation bellow at low
temperatures is exactly soluble and the partition function can be
presented in a closed form. Changing the atomic pseudo-spin
operators by a linear combination of Grassmann Fermi field yields
the fermionic generalized Dicke model with intensity-dependent
coupling defined by the Hamiltonian
\begin{eqnarray}
H&=&\sum_{j=1}^{N}\,\frac{\Omega}{2}\,\sigma_{(j)}^z+
\omega_{0}\,b^{\dagger}\,b+
\nonumber\\
&&\frac{g_{1}}{\sqrt{N}}\sum_{j=1}^{N}\,
\Bigl(b\,(b^{\dagger}\,b)^{\frac{1}{\,2}}
\alpha_{i}^{\dagger}\beta_{i}+b^{\dagger}\,
(b^{\dagger}\,b)^{\frac{1}{\,2}}\beta_{i}^{\dagger}\alpha_{i}\Bigr)+
\frac{g_{2}}{\sqrt{N}}\sum_{j=1}^{N}\,\Bigl(
b\,\beta_{i}^{\dagger}\alpha_{i} + b^{\dagger}\,
\alpha_{i}^{\dagger}\beta_{i} \Bigr)\, , \label{a28}
\end{eqnarray}
where again we are introducing different coupling between the
single mode bosonic field and the reservoir, $g_{1}$ and $g_{2}$
for rotating and counter-rotating terms, respectively. The
difference of the Hamiltonian for this model, given by Eq.
(\ref{a28}), when contrasted with the Hamiltonian of the
generalized Dicke model, Eq. (\ref{66a}), is the term
$(b^{\dagger}\,b)^{\frac{1}{\,2}}$ in the part of the interaction
Hamiltonian which generates the virtual processes. The fermion
integration on the generalized Dicke model is repeated for this
case, so we can take expressions given by Eq. (\ref{81}) and Eq.
(\ref{81a}) where the matrix $A$ is given by
\begin{equation}
A\,=\,\left( \begin{array}{cc} 0 & B\\ -C & 0
\end{array} \right)\, .
\label{83}
\end{equation}
In the equation above the quantities $B$ and $C$ are matrices with
components given now by
\begin{eqnarray}
B_{p\, q}&=&-\Biggl(\frac{\pi}{N\,\beta}\Biggr)^{\frac{1}{2}}\,
\Biggl(ip+\frac{\Omega}{2}
\Biggr)^{-\frac{1}{2}}\,\Biggl(iq-\frac{\Omega}{2}
\Biggr)^{-\frac{1}{2}}\nonumber\\
&&\times\left(\,g_1\,\Biggl(\frac{\pi}{\beta\omega_0}\Biggr)^{\frac{1}{2}}
\,\Bigl(b^*(0)\,b(0)\Bigr)^{\frac{1}{2}}\,\frac{b^{*}\,(q-p)}{\sqrt{\omega_{0}-i(q-p)}}
+g_{\,2}\,\frac{\,b\,(p-q)}{\sqrt{\omega_{0}-i(p-q)}}\right)
\label{83a}
\end{eqnarray}
and
\begin{eqnarray}
C_{p\, q}&=&\Biggl(\frac{\pi}{N\,\beta}\Biggr)^{\frac{1}{2}}\,
\Biggl(ip-\frac{\Omega}{2}
\Biggr)^{-\frac{1}{2}}\,\Biggl(iq+\frac{\Omega}{2}
\Biggr)^{-\frac{1}{2}}\nonumber\\
&&\times\left(\,g_1\,\Biggl(\frac{\pi}{\beta\omega_0}\Biggr)^{\frac{1}{2}}
\,\Bigl(b^*(0)\,b(0)\Bigr)^{\frac{1}{2}}\,\frac{b\,(p-q)}{\sqrt{\omega_{0}-i(p-q)}}
+g_{\,2}\,\frac{\,b^*\,(q-p)}{\sqrt{\omega_{0}-i(q-p)}}\right)\, .
\label{83b}
\end{eqnarray}
For this case, in order to find an approximate expression for the
partition function in the thermodynamic limit
($N\rightarrow\infty$), we use the approximation
$\det(I+A)=\det(\,I+BC)\simeq \exp(\,tr BC)$. Similar procedure
was used for the generalized Dicke model case. Then we can arrive
at the following approximate expression
\begin{eqnarray}
\frac{Z}{Z_{0}}\,&=&\,\int\,[d\eta(b)]\,\exp\Biggl
(-\,\pi\,\sum_{\omega}\,\Bigl(1-c(\omega)\Bigr)\,b^{\,*}
(\omega)\,b(\omega)\nonumber\\
&&+\,\,\pi\,\sum_{\omega}\,a(\omega)\,b^{\,*}
(0)\,b(0)\,b^{\,*}(\omega)\,b(\omega)+\,\pi\,\sum_{\omega}\,d(\omega)\,\Bigl(b^{\,*}
(0)\,b(0)\Bigr)^{\frac{1}{2}}
\nonumber\\
&&\times\biggl(\,b(\omega)\,b(\,-\,\omega)\,
+\,b^{\,*}(\omega)\,b^{\,*}(-\,\omega)\,\biggr)\,\Biggr)\,,
\label{zetainten}
\end{eqnarray}
where $a(\omega)$ and $c(\omega)$ of above equation are given,
respectively, by
\begin{equation}
a(\omega)\,=\frac{\pi\,g_{\,1}^{\,2}}{\beta\,\omega_0}
\tanh\biggl(\,\frac{\beta\,\Omega}{4}\,\biggr) \frac{1}{(\Omega
-i\omega)\,(\omega_{0}\,-\,i\,\omega)}\,, \label{term1}
\end{equation}
\begin{equation}
c(\omega)\,=\,g_{\,2}^{\,2}\,
\tanh\biggl(\,\frac{\beta\,\Omega}{4}\,\biggr) \frac{1}{(\Omega
+i\omega)\,(\omega_{0}\,-\,i\,\omega)}, \label{term2}
\end{equation}
and
\begin{equation}
d(\omega)\,=\,g_{\,1}\,g_{\,2}\,\Biggl(\frac{\pi}{\beta\,\omega_0}\Biggr)^{\frac{1}{2}}\,
\tanh\biggl(\,\frac{\beta\,\Omega}{4}\,\biggr) \frac{1}{(\Omega
+i\omega)\,(\omega^2_0\,+\,\omega^2)^{\frac{1}{2}}}\,.
\label{term3}
\end{equation}
In order to perform the functional integral given by Eq.
(\ref{zetainten}) we have separated the zero and non-zero modes,
so we can write
\begin{eqnarray}
\frac{Z}{Z_{0}}\,&=&\,\int\,db(0)\,db^*(0)\,e^{\pi\left(\,a(0)\,\bigl(b^*(0)\bigr)^2
b^2(0)\,+\,(c(0)-1)\,b(0)\,b^*(0)\right)}
\nonumber\\
&&\int\,\prod_{\omega\neq
0}db(\omega)\,db^*(\omega)\,e^{-\,\pi\,\sum_{\omega\neq
0}\,(1-c(\omega))\,b^*(\omega)\,b(\omega)}\nonumber\\
&&\times\,e^{\,\pi\,\sum_{\omega\neq 0}\,a(\omega)\,b^*
(0)\,b(0)\,b^* (\omega)\,b(\omega)\,+\,d(\omega)\,\bigl(b^{\,*}
(0)\,b(0)\bigr)^{\frac{1}{2}}\,\bigl(\,b(\omega)\,b(\,-\,\omega)\,+\,
b^*(\omega)\,b^*(-\,\omega)\,\bigr)}\,. \label{zetainten1}
\end{eqnarray}
It is quite difficult to compute in a close form the quantity
$\frac{Z}{Z_0}$ since the integrals that appear in Eq.
(\ref{zetainten1}) are not Gaussian. Nevertheless, at low
temperature ($\beta\rightarrow\infty$), we can make it evident
that the contribution coming from the counter-rotating terms
dominates over the rotating ones. We can take this limit in the
Eq. (\ref{term1}), Eq. (\ref{term2}) and Eq. (\ref{term3}) so we
get
\begin{eqnarray}
\lim_{\beta\rightarrow\infty}\,\frac{Z}{Z_{0}}\,=\int\,\prod_{\omega}
db(\omega)\,db^*(\omega)\,e^{-\,\pi\,\sum_{\omega}\,(1-c(\omega))\,b^*(\omega)\,b(\omega)}.
\label{zetaintenlimit}
\end{eqnarray}
Performing this last integral we have
\begin{eqnarray}
\lim_{\beta\rightarrow\infty}\,
\frac{Z}{Z_{\,0}}=\prod_{\omega}\,\biggl(1\,-\,c(\omega)\,\biggr)^{\,-\,1}\,,
\label{905}
\end{eqnarray}
where the term $c(\omega)$ in the limit $\beta\rightarrow\infty$
is
\begin{equation}
\lim_{\beta\rightarrow\infty}\,c(\omega)\,=\,\frac{\,g_{\,2}^{\,2}}{(\Omega
+i\omega)\,(\omega_{0}\,-\,i\,\omega)}\,. \label{term21}
\end{equation}
From this last equation we can see that the ratio $Z/Z_0$ is
non-analytic for $g_{\,2}=(\omega_0\,\Omega)^{\frac{1}{2}}$, thus
corresponding to a quantum phase transition. Therefore, in the
model at some range of the physical parameters, the
counter-rotating terms are dominant, leading to a phase
transition. Since we are not able to evaluate the ratio
$\frac{Z}{Z_{0}}$ given by Eq. (\ref{zetainten1}) for all
temperatures, it is not possible to describe the phase diagram of
the system near the quantum phase transition. The possibility that
appears a line of second-order phase transition for $\beta^{-1} >
0$ terminating at the quantum critical point must be investigated.

\section{Conclusions}

\quad $\,\,$In the present paper we are considering the issue of
the formation of the condensate with a superradiant phase
transition in spin-boson models with a single bosonic mode with
quite particular couplings between the single bosonic mode and a
environment of $N$ two-level atoms. We have assumed that the whole
system is in thermal equilibrium with a reservoir at temperature
$\beta^{-1}$. The interaction Hamiltonian of the models also
generate virtual processes. We first investigate a model based in
one discussed by Chang and Chakravarty, Legget and others. Second,
with the couplings $g_{1}$ and $g_{2}$ for rotating and
counter-rotating terms, respectively, we has defined the
generalized Dicke model. Finally we study the generalized Dicke
model with an intensity-dependent coupling. In this last model it
is necessary to use perturbation theory to investigate the
thermodynamic of the model. Nevertheless in the low temperature
limit the model is exactly soluble.

Studying the case where in identical two-level atoms act as a
thermal reservoir $(N \rightarrow \infty)$, we investigated the
thermodynamics of the three models using the path integral
approach with functional integration method. For generic $g^{2}$,
we found that the first model is unrealistic from the
thermodynamical point of view, since the entropy becomes negative
at some temperature. As we discussed, there are two different
solutions for this problem. The first one is an \textit{ad hoc}
procedure, disregarding the zero-mode, that can make the model
admissible. Another possibility is to assume that $g^{2}$ is a
bounded quantity. After any of these procedures, it is shown that
the partition function is analytic for all temperatures,
consequently there is no phase transition in the model. In the
second model the situation is more interesting. We studied the
nonanalytic behavior of thermodynamical quantities of the
generalized Dicke model by evaluating the critical transition
temperature and presenting the spectrum of the collective bosonic
excitations, for the case $g_{1}\neq 0$ and $g_{2}=0$, $g_{1}=0$
and $g_{2}\neq 0$ and also in the general case. Our result show
that it is possible to have a condensate with superradiance in a
system of $N$ two-level atoms coupled with one mode of a bosonic
field where only virtual processes contribute. It is important to
realize that the energy of the mode, which is not the
Nambu-Goldstone mode in Eq. (\ref{109}), is always larger than the
one in Eq. (\ref{111}), i.e., in the system where the condensate
appears due to the virtual processes. Therefore both processes,
real and virtual ones give different contributions to generate the
condensate. In the generalized Dicke model with the
intensity-dependent coupling we found a superradiant phase
transition at zero temperature, i.e., a quantum phase transition.

One interesting aspect of superradiance is the fact that initial
quantum fluctuations that are able to cause spontaneous emission
in a few atoms are amplified leading to fluctuations on a
macroscopic scale. We would like to point out that another
mechanism to enhance the importance of virtual processes, in
quantum field theory in the presence of macroscopic boundaries,
was proposed by Ford \cite{foc1}. In a local calculation, Ford
presented a mechanism based on the fact that the contribution of
various part of the frequency spectrum of the Casimir effect is an
oscillatory function. The contributions of different ranges of
frequency almost, but not quite, cancel out one another, and there
is the possibility of enhancing the magnitude of the effect by
altering the reflectivity of the boundary in selected frequency
ranges. A quite different mechanism for amplification of vacuum
fluctuations was proposed by Ford and Svaiter \cite{foc2}
\cite{foc3} using parabolic mirrors to produce large vacuum
fluctuations near mirror's focus. These authors studied the
renormalized vacuum fluctuations associated with a scalar and
electromagnetic field near the focus of a parabolic mirror. Using
the geometric optics approximation they found that the mirror
geometry can produce large vacuum fluctuations near the focus,
similar to what happens in the classical focusing effect by the
parabolic mirror geometry.

With the development of quantum information and its application in
computation and communication the entangled states have been
attracting enormous interest, since several quantum protocols can
be realized exclusively with the help of entangled states. In the
spin-boson model with the dipole-dipole interaction included, one
can address the issue of how does the dipole-dipole interaction
change the critical temperature, where again the system exhibits a
phase transition from the fluorescent to the superradiant phase.
An important subject for the near future is to study the degree of
entanglement between the $N$ two-level atoms and the bosonic mode
near the quantum phase transition in the generalized Dicke model
\cite{ent1} \cite{ent2} \cite{ent3}. This subject is under
investigation by the authors.

\section{Acknowlegements}

We would like to thanks G. Flores Hidalgo and H. J. Mosqueira
Cuestas for many helpful discussions. This paper was supported by
Conselho Nacional de Desenvolvimento Cientifico e Tecnol{\'o}gico
do Brazil (CNPq).

\end{document}